\documentclass[sigconf]{acmart}
\pdfoutput=1 
\usepackage{hyperref}
\PassOptionsToPackage{unicode}{hyperref}
\PassOptionsToPackage{naturalnames}{hyperref}
\usepackage{amsmath,amsfonts}
\usepackage{graphicx}
\usepackage{textcomp}
\usepackage{xcolor}
\usepackage{mathtools}
\DeclarePairedDelimiter\abs{\lvert}{\rvert}%
\usepackage[linesnumbered,algoruled]{algorithm2e}
\usepackage{algpseudocode}
\usepackage{amsthm}
\usepackage{comment}
\usepackage{bm}
\usepackage{subfig}
\newtheorem{definition}{Definition}
\usepackage{pgfplots}
\usepackage{pgfplotstable}
\usetikzlibrary{patterns, arrows, pgfplots.groupplots,hobby}
\usetikzlibrary{patterns}
\usetikzlibrary{positioning}
\usetikzlibrary{pgfplots.groupplots}
\usetikzlibrary{plotmarks}
\usetikzlibrary{calc}
\usepackage{booktabs}    
\usepackage{tabularx}
\usepackage{multirow}
\newcolumntype{L}[1]{>{\raggedright\let\newline\\\arraybackslash\hspace{0pt}}m{#1}}
\newcolumntype{C}[1]{>{\centering\let\newline\\\arraybackslash\hspace{0pt}}m{#1}}
\newcolumntype{R}[1]{>{\raggedleft\let\newline\\\arraybackslash\hspace{0pt}}m{#1}}
\definecolor{blue(pigment)}{rgb}{0.2, 0.2, 0.6}

\newcommand{\exactalg}{ERA}
\newcommand{\heuristicalg}{HRA}

\MakeRobust{\Call}

\hypersetup{draft}
\SetKw{Or}{\hspace{\algoskipindent}\itshape or\;}
\SetKw{And}{\hspace{\algoskipindent}\itshape and\;}

\AtBeginDocument{%
  \providecommand\BibTeX{{%
    \normalfont B\kern-0.5em{\scshape i\kern-0.25em b}\kern-0.8em\TeX}}}

\usepackage[all]{background}
\usepackage{stackengine}
\setstackEOL{\\}
\setstackgap{L}{\normalbaselineskip}
\SetBgContents{\color{gray}{\scriptsize \Longstack{PREPRINT - accepted at 59th ACM/IEEE Design Automation Conference (DAC) 2022}}}
\SetBgPosition{4.5cm,1cm}
\SetBgOpacity{1.0}
\SetBgAngle{0}
\SetBgScale{1.8}

\setcopyright{none}
\copyrightyear{2022}
\acmYear{2022}
\acmDOI{}

\acmConference[DAC'22]{Design Automation Conference}{July 10--14}{San Francisco, CA, USA}
\acmBooktitle{}
\acmPrice{15.00}
\acmISBN{}

\begin{document}
\title{Designing ML-Resilient Locking at Register-Transfer Level}


\author{Dominik Sisejkovic\textsuperscript{1}, Luca Collini\textsuperscript{2}\textsuperscript{*}, Benjamin Tan\textsuperscript{3}, Christian Pilato\textsuperscript{4},  \\Ramesh Karri\textsuperscript{2}, and Rainer Leupers\textsuperscript{1}}
\affiliation{\normalsize
	\institution{\textsuperscript{1} RWTH Aachen University, Germany, \textsuperscript{2} New York University, USA, \textsuperscript{3} University of Calgary, Canada, \textsuperscript{4} Politecnico di Milano, Italy}
	\streetaddress{}
	\city{} 
	\state{\{sisejkovic, leupers\}@ice.rwth-aachen.de, \{lc4976, rkarri\}@nyu.edu, benjamin.tan1@ucalgary.ca, christian.pilato@polimi.it\vspace{0.2in}} 
			\postcode{}
		\country{}
}








\renewcommand{\shortauthors}{}

\begin{abstract}
Various logic-locking schemes have been proposed to protect hardware from intellectual property piracy and malicious design modifications. Since traditional locking techniques are applied on the gate-level netlist after logic synthesis, they have no semantic knowledge of the design function. Data-driven, machine-learning (ML) attacks can uncover the design flaws within gate-level locking. Recent proposals on register-transfer level (RTL) locking have access to semantic hardware information. We investigate the resilience of ASSURE, a state-of-the-art RTL locking method, against ML attacks.
We used the lessons learned to derive two ML-resilient RTL locking schemes built to reinforce ASSURE locking. We developed ML-driven security metrics to evaluate the schemes against an RTL adaptation of the state-of-the-art, ML-based SnapShot attack. 
\end{abstract}




\maketitle
{
\footnotetext[1]{This work was done while L. Collini was with Politecnico di Milano, Italy.}
}


\section{Introduction}

Integrated circuits (ICs) are a critical layer for security in modern electronic systems. However, there are security concerns due to third parties in the supply chain. As external design houses and foundries have full access to the IC intellectual property (IP) during production, attackers could  reverse engineer the IP for malicious purposes, such as IP theft and hardware Trojan insertion~\cite{10.1145/3060403.3060495}. Design-for-trust methodologies aim to counteract such threats. Logic locking has been recognized as a premier technique to safeguard ICs throughout the supply chain~\cite{10.1145/3342099}. 
Logic locking builds on the concept of design obfuscation~\cite{yasin2020trustworthy, EPIC2008, 7362173, 8203496,tan2020benchmarking}, where designers insert key-driven logic to functionally and structurally alter ICs, thus concealing functional intent. Only the correct activation key unlocks the  intended functionality of the IC.

Recently, machine learning (ML) techniques have challenged the security of gate-level locking~\cite{sisejkovicVLSISoC21Survey}. ML-driven attacks exploit the predictable relation between the key value and the functional or structural aspects of locking. This has led to potent attacks that can either predict the correct key value or remove the locking circuitry from the netlist~\cite{OMLA2021,gnnunlock2021, sisejkovicJETCSnapShot2021, sailAttack2021}. While ML-driven attacks often lack output certainty, their applicability adds another requirement for logic locking success---\textit{prevention of key-related residue within the locking mechanics}. As long as the structural change is related to key values, it is possible to use ML to guess the keys. 
\begin{figure}[t]
	\centering
	\includegraphics[width=0.9\linewidth]{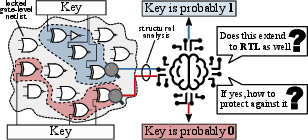}
	\caption{Machine learning vs. logic locking: impact on RTL?}
	\label{fig:motivation}
\end{figure}

Traditional gate-level locking schemes are limited to local changes and do not use the semantic information of the circuit, as logic synthesis and optimization disperse the semantics to a low granularity. Therefore, gate-level locking schemes operate ``blindly'' on the design without considering its functional traits. In response, RTL locking has emerged as a way to overcome this issue~\cite{pilatoTVLSI2020assure, 5401214}. At the RTL, locking can use the full spectrum of semantic information, including operations, constants, and control flow constructs. Hence, RTL locking is a promising basis to build ML-resilient locking. 
However, compared to gate-level locking, ML attacks on RTL locking remain unexplored as shown in Fig.~\ref{fig:motivation}. 

\textbf{Contributions:} Thus, this study explores ML resilience of RTL locking focusing on \textit{operation obfuscation}, where we: 
\begin{itemize}
	\item Introduce theoretical concepts to evaluate ML-resilience of RTL locking. 
	\item Expose security faults in ASSURE RTL locking~\cite{pilatoTVLSI2020assure}.
	\item Define ML-resilience security metrics for RTL locking.
	\item Introduce two ML-resilient locking algorithms: (1) \exactalg{}: \textbf{E}xact ML-\textbf{R}esilient \textbf{A}lgorithm and (2) \heuristicalg{}: \textbf{H}euristic ML-\textbf{R}esilient \textbf{A}lgorithm. 
	\item Evaluate  the locking algorithms against an RTL adaptation of the ML-based SnapShot attack.
\end{itemize}
To the best of our knowledge, the presented concepts and locking procedure are the first to address the challenges of ML resilience on RTL. \textbf{The implementation of this study will be made available to the community once published.}


\section{Background}

\subsection{Threat Model}
Since ML-driven attacks do not need a working chip (oracle) to succeed~\cite{sisejkovicJETCSnapShot2021}, our threat model includes the following assumptions. (1) The attacker has only access to the locked design in the form of a locked gate-level netlist or GDSII layout file. As a working copy of the locked chip is not available, this attack model is often referred to as \textit{oracle-less} (OL). Starting from the provided design level, the attacker can perform reverse engineering to recover the RTL design. To simulate the \textit{best-case scenario for the attacker}, we assume the attacker can retrieve an \textit{exact copy} of the initial, locked RTL design. (2) The attacker is aware of the algorithmic details of the applied locking scheme. (3) The location of the key inputs (pins) is known (i.e., \textit{distinct ambiguity}~\cite{10.1145/3342099}). In the rest of this study, we refer to the locked RTL design under attack as the \textit{target}.


\subsection{ML-Driven Structural Attacks}\label{ml-driven-struc-attacks}
\begin{figure}[t]
	\centering
	\includegraphics[width=\linewidth]{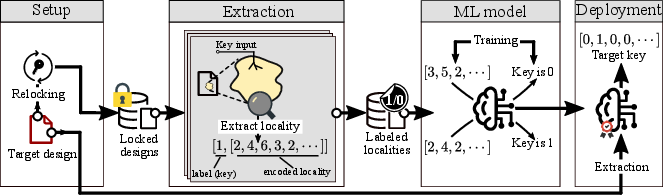}
		\vspace{-0.28in}
	\caption{ML-driven SnapShot attack flow.}
	\label{fig:snapshot}
		\vspace{-0.1in}
\end{figure}

In the OL model, an attacker has only access to the target design without I/O patterns. An ML-based attack has to exploit \textit{structural} key-related patterns to produce a (correct) key prediction. Thus, we selected the OL SnapShot attack~\cite{sisejkovicJETCSnapShot2021} as a basis for the evaluation. SnapShot was initially designed to attack locked gate-level netlists, following four major steps (Fig.~\ref{fig:snapshot}). First, the attack prepares a set of locked samples by \textit{relocking} (self-referencing) the target benchmarks with new keys. Second, a training set is assembled by extracting a netlist sub-graph for each single-bit key input from all data samples. The extracted sub-graphs are transformed into a vector of numbers, where each entry encodes a single gate from the derived sub-graphs. These vectors are referred to as \textit{localities}. In essence, \textit{a locality represents a key-affected portion of the netlist.} Next, the attack trains a dedicated ML model to associate localities with their respective key values. Finally, the trained ML model is deployed to predict the key of the target design. Since SnapShot has previously only been applied on gate level, we adjust the extraction and ML model of SnapShot to support RTL locking in this work.

Alongside SnapShot, the most prominent OL, ML-based attacks on gate-level locking are OMLA~\cite{OMLA2021}, GNNUnlock~\cite{gnnunlock2021}, and SAIL~\cite{sailAttack2021}. OMLA and GNNUnlock use graph neural networks, thus relying on a graph representation of the input design that is natural to gate-level netlists. SAIL exploits the deterministic and local changes of gate-level locking by learning to reverse the transformations induced by logic synthesis. Since we operate on RTL and assume a perfect reconstruction of the locked RTL, SAIL is not considered.

\subsection{The Concept of RTL Locking}
To protect the design, we focus on the locking techniques proposed in ASSURE~\cite{pilatoTVLSI2020assure}---one of the latest RTL locking policies. ASSURE offers three locking techniques: constant, branch, and operation obfuscation. Constant obfuscation extracts constants into the activation key. For example, \texttt{a = 4'b1101} is locked as \texttt{a = $K$}, where \texttt{$K$} is the 4-bit constant stored as the key. Branch locking works by XOR-ing the condition of the branch with a key bit, thereby inverting the condition if the value is 1. For example, the condition \texttt{a $>$ b} is locked as \texttt{(a <= b)$\wedge{} K$}. Operation locking works by inserting a key-controlled multiplexer to choose between a real and dummy operation. For example, the expression \texttt{a = b + c} can be locked either as \texttt{a = $K$ ? (b + c) : (b - c)} or \texttt{a = $K$ ? (b - c) : (b + c)}, depending on the value of \texttt{$K$}. In terms of security, constant obfuscation does not offer any apparent attack vectors, as the secret is fully omitted from the attacker. Branch obfuscation only affects the existing control flow based on the key, without inserting additional logic. Operation obfuscation manipulates the design by inserting additional logic \textit{depending on the existing one}. This dependence offers the potential for an attack. Therefore, \textit{we focus on the security of operation obfuscation}.  
\begin{figure}[t]
	\centering
	\includegraphics[width=\linewidth]{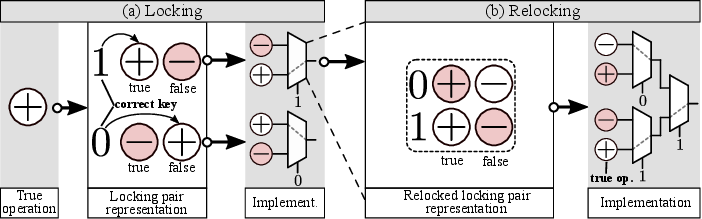}
		\vspace{-0.28in}
	\caption{ASSURE operation locking and representation.}
	\label{fig:assure-example}
		\vspace{-0.2in}
\end{figure}

\begin{figure*}[ht]
	\centering
	\includegraphics[width=0.95\linewidth]{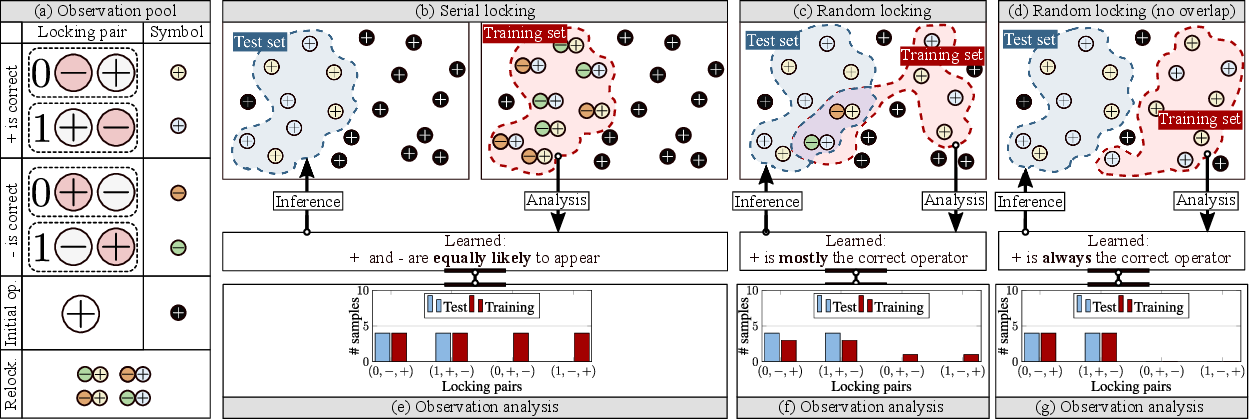}
		\vspace{-0.12in}
	\caption{Impact of operation selection on learning resilience in RTL locking.}
	\label{fig:ml-analysis}
		\vspace{-0.08in}
\end{figure*}
\textbf{Operation obfuscation:}~The security of this locking concept lies in the assumption that the attacker cannot guess which operation of the observed pair is the correct one. The paired real and dummy operations are called \textit{locking pairs}. In the general case, locking pairs are defined as $(T, T')$, where $T$ and $T'$ are the real and dummy operations, respectively. On RTL, a locking pair is typically implemented in the form of a \textit{ternary operator}. For instance, as depicted in Fig.~\ref{fig:assure-example}a, the real operation $+$ can be locked in the form of $(+, -)$ for the correct key value 1, or in the form of $(-, +)$ for the correct key value 0. Hence, an addition is always locked in pair with a subtraction, and vice versa. Note that all other operations have predefined locking pairs~\cite{pilatoTVLSI2020assure}. In case an already locked pair is relocked, both $T$ and $T'$ are locked separately. As shown in Fig.~\ref{fig:assure-example}b, relocking results in a tree of multiplexers, i.e., nested ternary operators. The compact notation of locked pairs from Fig.~\ref{fig:assure-example} is used in the rest of this study for visualization purposes.

\section{Learning Resilience for RTL Locking}\label{learning-res-for-rtl}

Learning attacks make predictions about the key by studying the locked design. 
A scheme that is secure against learning attacks is considered \textit{learning-resilient}~\cite{sisejkovicTCADDeceptive2021}. As discussed in~\cite{sisejkovicTCADDeceptive2021}, netlists that exhibit regular, repetitive structures \textit{maximize} the exposure of a locking scheme's mechanics, making it is easy to identify potential leakage points. This is because key-related structural changes are more likely to be identified within repetitive constructs.
To evaluate RTL locking for structural leakage, let us consider its workings on a structurally regular design that only contains connected $+$ operations.
The following challenge arises: \textit{how do we lock it without suggesting anything about the correctness of the key?} Let us consider the two methods of operation selection in ASSURE: \textit{serial} and \textit{random}. 
Furthermore, we need to take two data sets into account: test and training. The test set consists of locking samples for which the key is \textit{unknown} and represents the design under attack. The training set comprises locking samples that are added in additional relocking rounds of the target with \textit{known} keys. This process is also known as self-referencing~\cite{sailAttack2021, sisejkovicJETCSnapShot2021}. 
The attacker uses the  training set to collect observations about \textit{the relation between the locking pairs and the key}. The ensuing discussion follows visualizations in Fig.~\ref{fig:ml-analysis}, where we consider different locking scenarios using locking pairs and  symbols from Fig.~\ref{fig:ml-analysis}a, and the $+$ operation network.

\textbf{Serial selection}~is the standard selection in ASSURE. As shown in Fig.~\ref{fig:ml-analysis}b, the initial locking (test set) selects  $+$ operations for locking to create locking pairs in the form of $(+, -)$ and $(-, +)$ (encoded with a single symbol for simplicity). This "serial" selection always selects the operations in a serial manner w.r.t. the design topology. 
Due to the serial selection, the subsequent locking rounds (training set) select the \textit{same} operations as the test set for relocking; already-locked $+$ operations are extended with additional locking pairs. In the example, the left operation is selected for key value 1 and the right for key value 0, according to the rule of ternary expressions. As portrayed in Fig.~\ref{fig:ml-analysis}e, both the $+$ and $-$ operations \textit{are equally related to the key value 0 and 1}, resulting in confusing observations. This suggests that ASSURE is, in principle, learning-resilient. However, this case arises only due to the deterministic order of selecting operations---it can easily be broken by either using a longer training key to ensure locking untouched operations during training or by randomizing the order of selection.
Hence, the standard ASSURE procedure \textit{is not secure} w.r.t. data-driven attacks.

\textbf{Random selection} is depicted in Fig.~\ref{fig:ml-analysis}c.
Here, the samples from the test and training set are likely to overlap only to some extent. Hence, the observations of the training set are contaminated by some \textit{contradictory} observations. By analyzing the training set (Fig.~\ref{fig:ml-analysis}f), one can learn that the $+$ operation \textit{is more likely to be the correct one}. The random selection might result in a favorable outcome for the attacker when training and test samples do not overlap (Fig.~\ref{fig:ml-analysis}d). In this scenario, all observations from the training set (Fig.~\ref{fig:ml-analysis}g) suggest that the $+$ operation \textit{is always the correct one}. 
This knowledge can be used to infer a correct key. 

\subsection{Observations}~
$(1)$ Learning resilience on RTL can  be achieved if the likelihood of any operation in a locking pair is equally related to key value 0 and 1. $(2)$ Operation selection impacts learning resilience. $(3)$ 
The initial distribution of operation types determines if learning-resilience is achievable. Evidently, the effectiveness of learning-resilient locking \textit{should not depend on circuit features}. Even if real-world designs are not represented by the $+$ network, focusing on this biased case ensures that the scheme offers security even in general cases.

Based on the above observations, we can conclude that \textit{learning resilience on RTL is achievable if the occurrence frequency of every operation within a locking pair is equal for all operations in the pairings}. This is the case if the design has the same number of $+$ and $-$ operations after locking. In that case, any selection procedure for training results in an equal number of contradictory observations. In the next section, we introduce two locking algorithms that use this rule for learning resilience on RTL. 

\subsection{ASSURE Leakage Points}
We analyzed the serial selection of ASSURE~\cite{pilatoTVLSI2020assure}, and the \textit{current pairing of operations is leaky} as operations are incorrectly paired. For example, ASSURE assumes these pairs: $(*, +)$, $(+,-)$, and $(-,+)$. Here, * is paired with a +, but + is also paired with -. Hence, if the locked pair $(*, +)$ is encountered, the attacker can infer * as the correct operation, as $(+, *)$ does not exist. Similarly, leakage exists for modulo, xor, power, and division. Thus, \textit{currently ASSURE can be broken by analyzing operation pairs}. Hence, every operation must exist as a real and dummy operation with the same pair, e.g., $(*, /)$ and $(/, *)$. This fix applies to all evaluations in this study.

\section{ML-resilient RTL Locking}
Based on the discussion, we introduce the following definition:
\begin{definition}\label{learning-resilience-def}
	An RTL design is \textbf{learning-resilient} w.r.t. operation locking if the number of operations of type $T$ is equal to the number of operations of type $T'$ for each locking pair for which at least one operation of type $T$ or $T'$ is locked. 
\end{definition}
If neither $T$ nor $T'$ are involved in locking, the locked design is learning-resilient even if the number of $T$ and $T'$ operations is not balanced. The reason is that, during training, locking ``untouched'' $T$ and $T'$ operations does not provide feedback for the target samples.
Henceforth, \textit{secure} refers to security in the context of Def.~\ref{learning-resilience-def}.
Next, we introduce two ML-resilient locking algorithms built on top of ASSURE: \exactalg{}: \textbf{E}xact ML-\textbf{R}esilient \textbf{A}lgorithm and \textbf{H}euristic ML-\textbf{R}esilient \textbf{A}lgorithm. \exactalg{} guarantees security w.r.t. Def.~\ref{learning-resilience-def}, but requires a large key budget. \heuristicalg{} trades-off key length with security, yielding less secure solutions if the key budget is limited.

\setlength{\textfloatsep}{8pt}
\begin{algorithm}[t] \footnotesize
	\DontPrintSemicolon 
	\KwIn{Locking type $T$, operation distribution table $ODT$, RTL design D, and pair mode P}
	\KwOut{Number of used bits}
	$n \gets 0$\tcp*{Initialize used bits var}
	$T' \gets \Call{GetPairType}{T}$\;
	$o_{i} \gets \Call{RndSelect}{D.ops[T]}$\tcp*{Select a T-type op.} 
	$o_{j} \gets \Call{RndSelect}{D.ops[T']}$\;
	\uIf{$ODT[T] > 0$ \textbf{and} $!P$}{ %
		$\Call{AddPair}{D, o_{i}, T'}$\tcp*{Add T’ node to $o_{i}$}
		$ODT[T] \gets ODT[T] - 1$\;
		$ODT[T'] \gets ODT[T'] + 1$\;
		$n  \gets n  + 1$\;
	}
	\uElseIf{$ODT[T] < 0$ \textbf{and} $!P$}{
		$\Call{AddPair}{D, o_{j}, T}$\tcp*{Add T node to $o_{j}$}
		$ODT[T] \gets ODT[T] + 1$\;
		$ODT[T'] \gets ODT[T'] - 1$\;
		$n  \gets n  + 1$\;
	}
	\Else{ 
		$\Call{AddPair}{D, o_{i}, T'}$\tcp*{Add T’ node to $o_{i}$}
		$\Call{AddPair}{D, o_{j}, T}$\tcp*{Add T node to $o_{j}$}
		$n  \gets n   + 2$\;
	}
	
	\Return{$n$}	\;
	\caption{\sc{Lock}}
	\label{alg:lockingstep}
\end{algorithm}

\textbf{Operation distribution:}~The first step in ERA and HRA is to analyze operation distribution in the input RTL. We store this information in an \textit{operation distribution table} ($ODT$). 
For each $T$, the table stores a number representing the difference between the distribution of $T$-type and the locking-pair $T'$-type operations.
Assuming the pair $(+,-)$, a design with 7 "$+$" and 5 "$-$"  has the following $ODT$ entries: $ODT[+] = +2$ and $ODT[-] = -2$. A positive (negative) $ODT$ value indicates that the operation type has more (less) operations than its locking-pair type. The $ODT$ entries can inform a \textit{secure} design by balancing the number of $T$ and $T'$ operations. 

\textbf{The locking step:}~Algorithm~\ref{alg:lockingstep} outlines \textsc{Lock}, the common locking step for \heuristicalg{} and \exactalg{}. 
For a selected type $T$, the RTL $D$ is locked following three cases. If $ODT[T]$ is positive (lines 6-9), pair a new $T'$-type operation with an existing $T$-type to reduce the excess of $T$. If $ODT[T]$ is negative (lines 11-14), pair a new $T$-type operation with an existing $T'$-type to reduce the deficiency of $T$. Otherwise (lines 16-18), pair new $T$- and $T'$-type operations with existing operations. 
This is used  by \textit{specific operation-selection algorithms} to derive \heuristicalg{} and \exactalg{}. Before describing the locking algorithms, we introduce a security metric for resilience w.r.t. Def.~\ref{learning-resilience-def}.

\subsection{Security Metric for Learning Resilience}\label{metric}
$ODT$ entries can be used as a vehicle to \textit{measure security} in the context of Def.~\ref{learning-resilience-def}. To design a metric that indicates how "far" a locked design is from the optimal distribution, let us consider the following notation. The content of $ODT$ in iteration $j$ of a selected locking algorithm can be represented as the vector $\bm{v_{j}} = [x_{0}, \ldots, x_{l-1}]$, where $l$ is the number of available locking pairs, and $x_{i}=\abs{ODT[T]}$. Note that $\abs{ODT[T]} \equiv \abs{ODT[T']}$. A secure solution is reached if all entries of $ODT=0$. Hence, the optimal distribution can be defined as $\bm{v_{o}}=[y_{0}, \ldots, y_{l-1}]$, where $y_{i}=0$ for $i\in{[0,l-1]}$. Using this notation, we can define the learning-resilience security metric as:
\begin{equation}
	M_{sec}= 100 \cdot{\left(1 - \frac{d_{e}(\bm{v_{j}},\bm{v_{o}})}{d_{e}(\bm{v_{i}},\bm{v_{o}})}\right)},
\end{equation}
where $d_{e}$ is a modified version of the Euclidean distance, $\bm{v_{i}}$ the initial distribution vector of the target design, $\bm{v_{o}}$ the optimal distribution vector, and $\bm{v_{j}}$ the distribution vector after the $j$-th locking iteration. Note that $M_{sec}\in{[0,100]}$, where the highest value indicates $\bm{v_{j}} \equiv \bm{v_{o}}$. In that case, all locking-pair operation types are equally represented within the locked design, disabling the ability of ML to learn from relocking (as discussed in Section~\ref{learning-res-for-rtl}). Furthermore, the formulation of the Euclidean distance was adjusted as presented in Algorithm~\ref{alg:e-distance}. For a selected $\bm{v_{o}}$, the algorithm allows the exclusion of selected $\abs{ODT}$ values from the calculation, enabling two metric variants: \textit{restricted} and  \textit{global} learning resilience.

\MakeRobust{\Call}
\begin{algorithm}[t] \footnotesize
	\DontPrintSemicolon 
	\KwIn{Current vector $\bm{v_{j}}$ and optimal vector $\bm{v_{o}}$}
	\KwOut{Distance}
	
	$s \gets 0$\tcp*{Initialize sum var}
	\For{$i \gets 0; i < \abs{\bm{v_{o}}}; i++$}{
		\tcc{Check if value should be considered}
		\uIf{$\bm{v_{o}} \neq $ 'x'}{ 
			$s \gets s + (\bm{v_{o}}[i] - \bm{v_{j}}[i])^{2}$\;
		}
	} 
	
	\Return{$\sqrt{s}$}	\;
	\caption{$d_{e}$: Modified Euclidean Distance}
	\label{alg:e-distance}
\end{algorithm}

\textbf{Global security metric} ($M_{sec}^{g}$) considers all $ODT$ entries to determine $d_{e}$, regardless of whether operations from a selected locking pair are affected by locking or not.  This metric is suitable to guide heuristics when it is not clear which operation types will be locked. Thus, $M_{sec}^{g}$ describes \textit{the potential for exploitation within a design.} Since $M_{sec}^{g}$ considers all $ODT$ values, $\bm{v_{o}}$ does not contain any 'x' values. Hence, $M_{sec}^{g}$ is monotonic. 

\textbf{Restricted  security metric} ($M_{sec}^{r}$) only considers $ODT$ entries in which either $T$ or $T'$ are affected by locking. The reason is that an ML model cannot learn from operations from a selected locking pair if neither $T$ nor $T'$ operations are locked. In this sense, $M_{sec}^{r}$ captures the security of the design when only considering locked operations, i.e., \textit{the actual exploitability of the design}. If a selected locking pair is included during a locking procedure, certain 'x' values in $\bm{v_{o}}$ are set to 0. Thus, $M_{sec}^{r}$ is not monotonic.

If all types in $ODT$ are affected by locking, $M_{sec}^{r}\equiv M_{sec}^{g}$. Furthermore, $M_{sec}^{r}=100$ does not imply $M_{sec}^{g} = 100$, since some operation types are not affected by locking. However, if $M_{sec}^{g} =100$ then $M_{sec}^{r} = 100$.
These metrics are used by the locking algorithms.

\MakeRobust{\Call}
\begin{algorithm}[t] \footnotesize
	\DontPrintSemicolon 
	\KwIn{Key budget $k_{b}$ and RTL design $D$}
	\KwOut{Locked RTL design}
	$\Call{LoadODT}{D}$\tcp*{Populate ODT}
	$n \gets 0$\tcp*{Initialize used bits var}
	$\Theta \gets \{(T_{1}, T'_{1}), \ldots, (T_{n}, T'_{n})\}$\tcp*{Valid locking pairs}
	\While{$n  < k_{b}$}{
		$\vartheta \gets \Call{RndSelect}{\Theta}$\tcp*{Select a pair}
		$T \gets \Call{RndSelect}{\vartheta}$\tcp*{Select a type}
		
		\While{$\abs{ODT[T]} > 0$}{
			$s\gets \Call{Lock}{T, ODT, D, False}$\tcp*{Apply lock (Algorithm 1)}
			$n \gets n + s$\;
		}
	}
	
	\Return{$D$}	\;
	\caption{\exactalg{}: Exact ML-Resilient Algorithm}
	\label{alg:exact-alg}
\end{algorithm}



\subsection{\textbf{E}xact ML-\textbf{R}esilient \textbf{A}lgorithm (\exactalg{})}
\exactalg{} (Algorithm~\ref{alg:exact-alg}) ensures that locking always yields a secure design even if the key budget is exceeded.
While the key budget is not exceeded (line 4), \exactalg{} randomly selects a type $T$ from a randomly selected pair $\vartheta$ from  valid locking pairs $\Theta$ (lines 5-6). To ensure a secure solution after each selection, the algorithm repeats the locking for the selected type until $ODT[T]$ reaches 0 (lines 7-10). 
This way the selected operation pairs yield a balanced solution.  Thus, $M_{sec}^{r}=100$ after each locking round even if the cost is more than allowed. Hence, \exactalg{} prioritizes security over cost.
\exactalg{} 
always locks all selected pairs until $ODT[T]$ reaches 0---all affected pairs are guaranteed to be balanced. The security evaluation of an \exactalg{}-locked design will result in $M_{sec}^{r}=100\%$, but not necessarily in $M_{sec}^{g}=100\%$. The former states that all affected pairs are perfectly balanced. The latter indicates that other parts of the design are exploitable by ML \textit{if not locked properly} (if $M_{sec}^{g}<100\%$).


\begin{algorithm}[t] \footnotesize
	\DontPrintSemicolon 
	\KwIn{Key budget $k_{b}$ and RTL design $D$}
	\KwOut{Locked RTL design}
	$\Call{LoadODT}{D}$\tcp*{Populate ODT}
	$n \gets 0$\tcp*{Initialize used bits var}
	$\bm{v_{i}} \gets \Call{ExtractVector}{D.ODT}$\tcp*{Initial vector}
	$\Theta \gets \{(T_{1}, T'_{1}), \ldots, (T_{n}, T'_{n})\}$\tcp*{Valid locking pairs}
	\While{$n  < k_{b}$}{
		$M_{sec}^{g} \gets 0$\tcp*{Track max metric value}
		$j \gets 0$\tcp*{Track operation index}
		$P \gets \Call{RndBoolean}{\null}$\tcp*{Include randomness}
		\uIf{$P$}{
			$j \gets \Call{RndSelect}{\abs{\Theta}}$\;
		}
		\Else{
			$\Call{Shuffle}{\Theta}$\;
			\For{$i \gets 0; i < \abs{\Theta}; i++$}{
				$\Call{Lock}{\Theta[i][0], ODT, D, False}$\;
				$\bm{v_{j}} \gets \Call{ExtractVector}{D.ODT}$\;
				$M_{i} \gets \Call{EvalMetric}{\bm{v_{i}}, \bm{v_{j}}}$\;
				$\Call{UndoLock}{D}$\tcp*{Undo last lock}
				\uIf{$M_{i} > M_{sec}^{g}$}{
					$M_{sec}^{g} \gets M_{i}$\;
					$j \gets i$\;
				}
			}
		}
		$s\gets \Call{Lock}{\Theta[j][0], ODT, D, P}$\tcp*{Apply lock (Algorithm 1)}
		$n \gets n + s$\;
	}
	
	\Return{$D$}	\;
	\caption{\heuristicalg{}: Heuristic ML-Resilient Algorithm}
	\label{alg:heuristic-alg}
\end{algorithm}

\subsection{\textbf{H}euristic ML-\textbf{R}esilient \textbf{A}lgorithm (\heuristicalg{})} 
HRA (Algorithm~\ref{alg:heuristic-alg}) performs iterative fine-grained balancing of locking-pairs in the target design to get closer to the secure solution at every step without exceeding the key budget. 
While key bits are available (line 5), \heuristicalg{} randomly chooses (line 8). Either a random operation type is chosen (line 10) or the best type is chosen (lines 12-21). The latter case evaluates all locking pairs in $\Theta$ and checks which one yields the \textit{highest} increase in $M_{sec}^{g}$. In both cases, the selected pair is locked by the \textsc{Lock} function (line 23).
As \heuristicalg{} performs fine-grained design adjustments, it uses the exact key budget and trades off against the guarantee to reach a secure solution. Since \heuristicalg{} ensures that every step increases security and decreases operation imbalances, it must be guided by the monotonic $M_{sec}^{g}$ metric.

\begin{figure}[t!]
	\centering
		\includegraphics[width=\columnwidth]{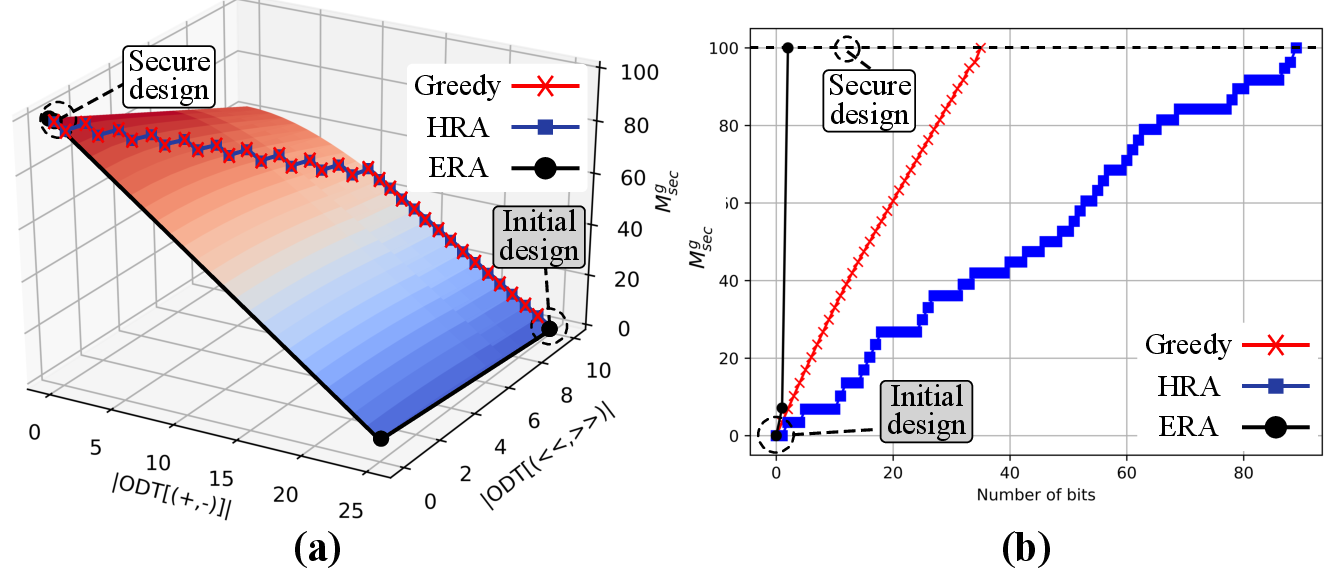}
			\vspace{-0.25in}
	\caption{Security metrics: (a) search space and (b) evolution.}
	\label{fig:metric}
\end{figure}
\begin{figure*}[ht]
	\centering
	\subfloat[KPA per benchmark]{
		\begin{tikzpicture}[scale=0.49]
			\pgfplotstableread[row sep=\\,col sep=&]{
				bench & v1 & v2 & v3 \\
				1 &    75.00 &	65.00 &	40.00 \\
				2 &    66.00 &	48 &	54.00 \\
				3 & 89.00 &	89.00 &	42.00\\
				4 &    66.00 &	53.00 &	49.00 \\
				5 & 83.00 &	85.00	& 57.00	\\
				6 & 68.00 & 58.00 & 51.00 \\
				7 & 53.00 & 54.00 & 59.00 \\
				8 & 99.00 & 100 &  47.00\\
				9 & 95.00 & 95.00 & 48.00 \\
				10 & 68.00 & 79.00 & 36.00 \\
				11 & 82.00 & 82.00 & 47.00 \\
				12 &   63.00 &	89.00 & 40.00\\
				13 &  90.00  0 &	100.00  &	51.00 \\
				14 &  50.00  0 &	50.00  &	50.00 \\
			}\dataset
			
			\begin{axis}[ybar,
				ylabel={KPA (\%)},
				xlabel={Benchmarks},
				ylabel style={at={(-0.001,0.5)},anchor=north},
				legend style={at={(0.1,0.96), font=\Large},
					anchor=north,legend cell align={left}},
				legend image post style={scale=1.4},
				enlarge x limits={abs=0.8cm},
				ymajorgrids = true,
				bar width=.38cm,
				width=30cm, height=4.3cm,
				ymin=0,ymax=140,
				ytick={0, 25, 50, 75, 100},
				xtick align=inside,
				xtick={1,2,3,4,5, 6, 7,8, 9, 10, 11,12,13,14},
				xticklabel style={align=center},
				xticklabels = {
					{\tt DES3},
					{\tt DFT},
					{\tt FIR},
					{\tt IDFT},
					{\tt IIR},
					{\tt MD5},
					{\tt RSA},
					{\tt SHA256},
					{\tt SASC},
					{\tt SIM\_SPI},
					{\tt USB\_PHY},
					{\tt I2C\_SL},
					{\tt N\_2046},  
					{\tt N\_1023},  
				},
				ylabel style ={font=\Large},
				xlabel style ={font=\Large},
				tick label style={font=\large},
				legend entries={ASSURE, HRA, ERA},
				legend columns=3,
				]
				\addplot[blue(pigment),thick, sharp plot,dashed, update limits=false,forget plot] 
				coordinates {(0,50) (15,50)} 
				node[above] at (axis cs:13.7,50) {\Large Random guess};
				\addplot[draw=black,fill=white] table[x=bench,y=v1] \dataset;
				\addplot[draw=black,fill=white, postaction={pattern=north east lines}] table[x=bench,y=v2] \dataset;
				\addplot[draw=black,fill=gray!80] table[x=bench,y=v3] \dataset;
				
			\end{axis}
			\label{fig:sim_ambient_temp}
		\end{tikzpicture}
	}   \subfloat[Average KPA]{
		\begin{tikzpicture}[scale=0.49]
			\begin{axis}[ybar,
				ylabel={KPA (\%)},
				xlabel={RTL locking algorithms},
				ylabel style={at={(-0.004,0.5)},anchor=north},
				legend style={at={(0.5,0.96), font=\Large},
					anchor=north,legend cell align={left}},
				legend image post style={scale=1.4},
				enlarge x limits={abs=1cm},
				ymajorgrids = true,
				bar width=.38cm,
				width=6.3cm, height=4.3cm,
				ymin=0,ymax=140,
				ytick={0, 25, 50, 75, 100},
				xtick align=inside,
				xtick={1,2,3},
				xticklabel style={align=center},
				xticklabels = {ASSURE, HRA, ERA},
				ylabel style ={font=\Large},
				xlabel style ={font=\Large},
				tick label style={font=\large},
				legend columns=3,
				every node near coord/.append style={
					anchor=north,
					yshift=4ex,
					xshift=-2ex,
					font=\Large,
					rotate=90
				},
				every node near coord/.append style={
					/pgf/number format/fixed, 
					/pgf/number format/fixed zerofill,
					/pgf/number format/precision=2
				},
				]
				\addplot[blue(pigment),thick, sharp plot,dashed, update limits=false,forget plot] 
				coordinates {(0,50) (4,50)} 
				node[above] at (axis cs:3,50) {};
				\addplot[draw=black,fill=white,nodes near coords]coordinates {(1.2,74.78)};
				\addplot[draw=black,fill=white,nodes near coords, postaction={pattern=north east lines}]coordinates {(2,74.26)};
				\addplot[draw=black, nodes near coords,fill=gray!80]coordinates {(2.7,47.92)};
			\end{axis}
			\label{fig:}
		\end{tikzpicture}
	}  
	\vspace{-0.15in}
	\caption{Evaluation results for the ML-based SnapShot attack on RTL locking.}
	\label{fig:eval}
	\vspace{-0.1in}
\end{figure*}
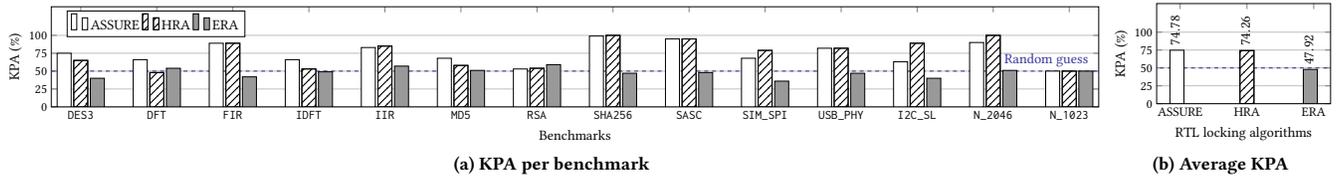

\subsection{Metric-Guided Design}
The proposed metrics can be used to design various locking algorithms targeting learning resilience. Let us consider a design with the following $ODT$ entries: $\abs{ODT[(+,-)]}=25$ and $\abs{ODT[(<<,>>)]}=10$. As depicted in Fig.~\ref{fig:metric}a, $M_{sec}^{g}$ represents a smooth, monotonic surface. 
Fig.~\ref{fig:metric}b presents the evolution of the metric in each step. The goal of locking is to move the target design from the initial point (bottom right) to the secure point (top left). The path between these two points represents different heuristic approaches. \exactalg{} forces selected $ODT$ values to 0, thus jumping in two steps to the secure solution alongside the edges of the surface. \heuristicalg{} travels in the steepest direction, taking small steps and remaining on the highest line across the surface. A greedy approach (same as \heuristicalg{} where $P$ in line 8 is always false) traverses the same points as \heuristicalg{}. Fig.~\ref{fig:metric}b suggests that a greedy approach is more efficient than HRA since it reaches full security (i.e., metric equal to 100) with fewer key bits. However, a greedy approach has a negative consequence: reversibility. An attacker can  reverse the locking procedure alongside the steepest decreasing direction. Therefore, including random locking decisions within \heuristicalg{} (variable $P$) thwarts reversibility---even if it takes longer to get to the secure solution. Similar observations hold for $M_{sec}^{r}$ since $M_{sec}^{r}\equiv M_{sec}^{g}$ when all $ODT$ entries are affected.

\section{Evaluation}
We evaluate ASSURE-based locking policies against the state-of-the-art ML-based SnapShot attack on a subset of the benchmarks used in~\cite{pilatoTVLSI2020assure}. 
Some benchmarks were excluded due to the low number of operations. We also composed two synthetic benchmarks: \texttt{N\_2046} and \texttt{N\_1023}, representing a fully imbalanced (biased) design (a network of 2046 $+$ operations) and a fully balanced design (a network with 1023 $+$ and $-$ operations), respectively. We consider ASSURE (serial implementation), \heuristicalg{}, and \exactalg{}. Note that the cost of the proposed algorithms are in line with the original ASSURE, as the cost of a locking pair per key bit has not changed~\cite{pilatoTVLSI2020assure}.

\textbf{SnapShot for RTL:} We adapted SnapShot (Fig.~\ref{fig:snapshot}) to learn RTL key leaks by extracting all key-controlled pairs $[K[i], C_{1}, C_{2}]$, where $K[i]$ is the key-bit value, and $C_{1}$, $C_{2}$  are encodings for an operation pair. We assign each type a unique integer. The extractor uses the Pyverilog library~\cite{Takamaeda:2015:ARC:Pyverilog}. 
Instead of one neural network type as in~\cite{sisejkovicJETCSnapShot2021}, we use auto-sklearn~\cite{feurer-neurips15a},  a library for automatic ML (auto-ml) model exploration. Auto-ml searches for an ML model among the implementations and optimizes the hyperparameters. 
We selected 600 seconds per attack iteration as this was enough for the attack to converge.

\textbf{Attack setup:}
The test set for each algorithm comprises every benchmark locked 10 times with different keys. We assembled the training set by \textit{relocking} each test sample 1,000 times with different keys. Relocking was performed with random ASSURE locking so that all parts of the design were used for learning; thus, simulating the most effective attack. Both test and training keys are set to 75\% of the operations for each benchmark. This was exceeded for the \texttt{N\_2046} benchmark, as its perfect imbalance requires a 100\% key budget for \exactalg{}. We assumed a best-case scenario for the attacker: a perfect reconstruction of the initial, locked RTL.

\textbf{Accuracy metric:} Key Prediction Accuracy (KPA) is used to measure attack success~\cite{sisejkovicJETCSnapShot2021}. N\% KPA indicates that N\% of the key bits are correctly predicted. A random guess has 50\% KPA.


\subsection{Results and Discussion}
\textbf{Results:} Fig.~\ref{fig:eval}a presents the KPA evaluation results per locking algorithm and benchmark, and Fig.~\ref{fig:eval}b presents the average KPA across all benchmarks. 
SnapShot correctly predicts 74.78\% key bits for the original ASSURE implementation, on average. The average KPA for \heuristicalg{} is slightly lower, 74.26\%. \exactalg{} averages $\sim$47.92\% KPA with consistent KPA values around (or lower) than a random guess.

\textbf{Lessons learned:} 
SnapShot's success on \heuristicalg{} is at first surprising since it is supposed to have a higher level of security than non-ML-driven serial locking. However, since we use a key budget of 75\% of the available operations, parts of the design remain unaffected by locking. Hence, the training step \textit{can extract knowledge} about the design for an educated guess ($\sim$24 percentage points better than random). Once all operations are fully balanced---as guaranteed by \exactalg{}---the training fails to extract useful observations. The above leads to a significant conclusion: \textit{\textbf{when it comes to ML-driven attacks, half measures are not effective.}} Data-driven approaches can exploit even the slightest imbalance. In contrast, half-way measures can mitigate non-ML-driven attacks, e.g., slightly increasing the key length can deteriorate a brute-force attack. 
While \heuristicalg{} appears less promising, the heuristic is useful if multiple security objectives must be reached, such as learning-resilience, output corruptibility, and Boolean Satisfiability (SAT)-resistance~\cite{SMT2020}. Since \exactalg{} makes coarse-grained modifications, it might create radical changes in the design. \heuristicalg{} improves learning resilience of locked designs alongside other objectives \textit{in smaller and controlled locking steps} as it only decreases operation imbalance.

\textbf{Limitations and opportunities:} This study exploits individual locking pairs---but is there a "global bias" among designs? If so, this bias could help determine the correct function of locked designs. The metric in Section~\ref{metric} can extract the initial distance for selected designs by considering the distance between the initial distribution and the optimal one. Are the locking algorithms resilient to oracle-guided attacks? Moreover, locking has recently been explored in combination with high-level synthesis~\cite{9586159, TAOLocking2018, 10.1145/3410337}. Future efforts should evaluate the problem of learning resilience on this abstraction level and address the mentioned challenges.

\section{Conclusion}
We introduced the first concepts on designing and evaluating RTL locking using ML-based attacks on operation obfuscation, and proposed two ML-resilient locking algorithms. The heuristic algorithm is a controlled procedure that decreases the imbalance of operations in an RTL design in small steps, adhering to the allowed key budget. The exact algorithm guarantees ML resilience but can exceed a key budget. We presented a security metric to assess resilience of RTL locking to ML attacks that can guide the design process of heuristic locking. Finally, we presented the first ML-based oracle-less attack on RTL locking by adapting the state-of-the-art SnapShot attack. 

\vspace{0.5em} \noindent \textbf{ACKNOWLEDGMENTS} \\
R. Karri was supported in part by ONR Award \# 
N00014-18-1-2058, NSF Grant \# 1526405, NYU Center for Cybersecurity, and NYUAD
Center for Cybersecurity. \vspace{-0.5em}

\bibliographystyle{ACM-Reference-Format}
\bibliography{bibliographyshort}

\end{document}